# DBBRBF- Convalesce optimization for software defect prediction problem using hybrid distribution base balance instance selection and radial basis Function classifier

**Mrutyunjaya Panda**
Reader, P.G. Department of Computer science and Applications
Utkal University, Vani Vihar, Bhubaneswar-4, Odisha, India
Mrutyunjaya74@gmail.com

**Abstract:** Software are becoming an indigenous part of human life with the rapid development of software engineering, demands the software to be most reliable. The reliability check can be done by efficient software testing methods using historical software prediction data for development of a quality software system. Machine Learning plays a vital role in optimizing the prediction of defect prone modules in real life software for its effectiveness. The software defect prediction data has class imbalance problem with low ratio of defective class to non-defective class, urges an efficient machine learning classification technique which otherwise degrades the performance of the classification. To alleviate this problem, this paper introduces a novel hybrid instance based classification by combining distribution base balance based instance selection and radial basis function neural network classifier model (DBBRBF) to obtain best prediction in comparison to the existing research. Class imbalanced data sets of NASA, Promise and Softlab were used for the experimental analysis. The experimental results in terms of Accuracy, F-measure, AUC, Recall, Precision and Balance show the effectiveness of the proposed approach. Finally, Statistical significance tests are carried out to understand the suitability of the proposed model.

## 1. Introduction

Knowing that Software defect may cause serious consequences in terms of huge financial and human losses in today's software intensive system, early detection of defect prone modules before release of any new software attracts lots of attention [1]. In spite of lot of research to obtain the quality software is ON, still the poor performance issues with reliability become a major concern for the researchers due to the inherent problems in anyone or all of the following: no clear understanding in the requirement, coding errors and insufficient software testing before release etc. to name a few [2]. It is also envisioned that the software defect prediction shall be carried out in entirety rather than investigating each individual component in isolation and further, the design choice shall be made judiciously in order to avoid the loss of generality and/or to avoid the useless results.

Since there are scarcity in getting quality data to have better software defect prediction model which are usually not only noisy but also suffers from class imbalance problem, data pre-processing followed by machine learning application are being proposed by many researchers [3, 4]. Researchers have also opined for a good machine learning model for classification defective and non defective software modules with a consideration that the cost involved in mis-classification of defective ones as non defective are more than the other ones, where the testing time is increased [5,6,7]

**Motivation:**

While solving the software defect prediction problem, incorporation of both labeled and unlabeled data in the machine learning process may lead to best possible classification results. To this end, many researchers have used Graph based learning with application of sparse theory on the dataset for pairwise relationship [8]; collaborative representation by the authors [9]; metrics- based [10]; class imbalance [11]; Dictionary learning [12], traditional methods like: Support Vector Machine (SVM) [13] , Naive Bayesian (NB) [14], Neural Network [15] and the list goes on. It is observed that performance of the traditional methods severely limited with respect to lack of common feature representation and selection of a good feature selection algorithm in order to deal with sparse nature of the software prediction dataset.
The bulk of defect prediction experiments based on the NASA Metrics Data Program data sets may have led to erroneous findings. This is mainly due to repeated data points potentially causing substantial amounts of training and testing data to be identical [16].
Thus it is concluded that by using feature selection techniques the time & space complexity for defect prediction reducing without effecting the prediction accuracy.
Future scope of project lies in performing double pre-processing of dataset by applying instance filtering along with attribute selection [17].
This motivates us to explore a novel distributed base balance method based instance selection to remove the duplicate instances of data followed by radial basis function classifier to improve the performance of software defect prediction model.

**Research Objective:**The main objective of this research is to develop a software defect prediction model that is not only most accurate and fast but also have a good balance. We propose a supervised Instance Selection filter Using distributed base balance (DBB) method to overcome the problems mentioned above by filtering out irrelevant instances. The proposed approach performs a classification with Radial basis function neural network classifier (RBF). To evaluate the proposed DBBRBF, this paper investigates the following research questions:

• RQ1: does DBBRBF generate performance practical to use compared with other models under the same way of learning?

• RQ2: can DBBRBF produce the prediction performance comparable to others related research available?

The prediction performance of DBBRBF is assessed through Kruskal-Wallis test with post-hoc test along with Mann-Whitney test for evaluation of the results. From all the experiments, it may be concluded that the proposed DBBRBF classifier is able to predict the software defects efficiently which will help in making better decision .
The remainder of this paper is organized as follows. In Section 2, we describe related work and then, our proposed instance based selection with classification model is explained in Section 3. While experimental setup is described in Section 4, the results of the experiments conducted are described in Section 5. Discussion of the results with possible threats to validity of our approach provided in Section 6. Finally, Conclusion and future scope is presented in Section 7.

## 2. Related work

The authors [18] propose a novel machine learning approach using multiple linear regression model to predict bug proneness in software defect prediction Eclipse JDT Core data. Considering Software defect prediction as a classification task, the authors proposes SMOTE (Synthetic Minority Over-sampling Technique) ensemble based approach to effectively deal with the class imbalance problem of the datasets used and to achieve high accuracy[19].

In [20], the authors propose to help software developers by identifying software defects basing on the existing software metrics with various classification techniques. It is proposed to evaluate software defect prediction via Maximal Information Coefficient with Hierarchical Agglomerative Clustering (MICHAC) method on 11 widely studied NASA projects using three different classifiers such as: Naive Bayes, RIPPER and Random Forest) with four performance metrics (precision, recall, F-measure, and AUC) and opines their effectiveness in comparison to others [21].

The authors [22] discusses the application of data mining in software defect prediction for both static and dynamic defects, clone defects etc and highlights its importance to assist in software engineering tasks. A good overview on the data quality of the NASA MDP datasets [24] is presented in [23] as reported in [25] where comprehensive rules for data cleansing are used for software defect prediction.

Six state-of-the-art within-project defect prediction approaches such as: naive Bayes, Decision tree, Logistic regression, K-nearest neighbor, random forest and Bayesian network are used by the authors [26] with 14 performance metrics applied to PROMISE repository [27] to check their effectiveness.

The authors [28] have investigated comparison of 37 classification algorithms over NASA datasets and concluded that Bagging shows a better performance than the rest of classifiers in fault detection systems.

## 3. Software defect Prediction datasets

This section discusses about the software defect prediction datasets used in this paper.

To carry out the empirical evaluation of the proposed technique, we selected two datasets from the PROMISE repository, which contains data made publicly available in order to encourage repeatable, verifiable, refutable, and/or improvable predictive models of software engineering [24].

NASA's Metrics Data Program Data Repository [24,31] is a database that stores problem, product, and metrics data. The primary goal of this data repository is to provide project data to the software community. In doing so, the Metrics Data Program collects artifacts from a large NASA dataset, generates metrics on the artifacts, and then generates reports that are made available to the public at no cost. The data that are made available to general users have been sanitized and authorized for publication through the Metrics Data Program Web site by officials representing the projects from which the data originated.

The tera-PROMISE Repository [25] is a research dataset repository specializing in software engineering research datasets. The datasets we employed are related to versions 4.0, 4.2 and 4.3 of the jEdit system[34], a well-known text editor written in Java as proposed in Chidamber and Kemerer (CK) metrics [29]. IDE is used for understanding the various characteristics of any software applications with true and false representation showing fault existence or not respectively [30].
Further, SOFTLab dataset versions of AR4, AR5 and AR6 [32,33] used for the experimental purpose in this paper. Softlab is the Software Research Laboratory in Bogazici University, Istanbul, Turkey. Mozilla4 data set [34,35] is used to to study the effect of class size on defect-proneness in the Mozilla product and main intention is to create a conditional Cox Proportional Hazards Model with the following features: (1) % id: A numeric identification assigned to each separate C++ class where it is to be noted that the id's do not increment from the first row to the last data row), (2) start: A time infinitesimally greater than the time of the modification that created this observation (practically, modification time). When a class is introduced to a system, a new observation is entered with start=0, (3) end: Either the time of the next modification, or the end of the observation period, or the time of deletion, whichever comes first. (4) event: event is set to 1 if a defect fix takes place at the time represented by 'end', or 0 otherwise. A class deletion is handled easily by entering a final observation whose event is set to 1 if the class is deleted for corrective maintenance, or 0 otherwise. (5) size: It is a time-dependent co-variate and its column carries the number of source Lines of Code of the C++ classes at time 'start'. Blank and comment lines are not counted and (6) state: Initially set to 0, and it becomes 1 after the class experiences an event, and remains at 1 thereafter. At last, to check whether the effect size is prone to the software defect.
The descriptive statistics of the metrics and fault data for the employed datasets are reported in Table 1, Tables 2 and Table 3.

## 4. Proposed Methodology

This section highlights the proposed methods adopted in the paper.

### 4.1. Distribution-based balancing of dataset

The Distributed based balancing of dataset (DBB) [37] is a supervised instance selection approach, which is proposed in this paper to address the challenging tasks of data imbalance problem in almost all software defect prediction datasets available. This process belongs to the category of data-based algorithms that combine under- and over-sampling with total replacement of the training set, so that fully balances all the classes available in the dataset. This is achieved by performing the following on the training dataset such as: (1) Over-samples classes with less than b documents, (2) Under-samples classes with more than b documents (3) Removes or reduces over-lapping among classes and finally, (4) Fully balances all classes. The Pseudo-code for the DBB process is presented in Figure 1.
In this paper, sampling from a Poisson distribution is used in an approximate way for reducing the sampling time and gets similar sampled values. For balancing the data imbalance, Gaussian Distribution is used to learn and re-sample new instances. At a time, 30 number of instances are considered to re-sample per class label. The reason behind using Gaussian distribution based balancing is its natural ability to build the model with correct approximation.

## 4.2. Radial Basis Function (RBF) Neural Network Classifier

Radial Basis Function Neural Network (RBFNN) [38] is a three layered feed-forward neural network consisting of linear inputs at first layer, second layer is called as hidden layer with non-linear Gaussian functions and the third layer linearly combines the Gaussian output to present the result. Radial basis function is used as activation function here. In RBFNN, the following parameters are considered for necessary optimization such as: weights update between hidden and output layer, activation function with its center and distribution of center of activation function and number of hidden neurons. Moore-Penrose generalized pseudo-inverse is used for weight updation in order to overcome the issues of the traditional neural network algorithms in terms of : update in learning rate, number of epochs, stopping criteria and local optima etc.. Moreover, it is faster in training the model with generalization capability. This tempted us to use this method for our proposed approach to have a good real life applications. The Gaussian kernel is used as radial basis function for activation function. The centers and distribution of activation functions uses K-Means clustering algorithm for its simplicity and efficient implementation. The choice of number of hidden layers relies on the universal approximation theory which dictates that a single hidden layer with a sufficient number of hidden neurons may approximate the activation function to any arbitrary level of accuracy. However, if one chooses the sufficient number of hidden neurons, then there is always a chance that the activation function with its center and distribution of center of activation function be non-deterministic. The Pseudo-code for RBFNN is presented in Figure 2.

# 5. Experimental setup

All the experiments are performed in Intel Core i5 , 1TB HDD, 2.64CPU with 4GB RAM in Windows 10 environment. The experimental setup is provided in Figure 3.

In this paper, we apply 10-fold cross validation to historical data extracted from NASA MDP, PROMISE repository in order to obtain training and testing dataset. In this, for each case of cross validation, the full dataset is partitioned into 10 subsets out of which 9 subsets are selected randomly for training and the rest one is used for testing the model proposed.

It is understood from the available literature that feature or attribute selection alone has not been able to perform well using any of the classifiers to the best possible way with highest accuracy, balance, F-score etc. Hence, we are tempted to see whether instance based selection instead of attribute selection cam address the issues raised by many researchers in this area while building an efficient software prediction model. To that effect, we use a novel Distribution based balancing (DBB) of dataset as a supervised instance selection with Poisson distribution for sampling the data and Gaussian distribution for balancing the data with 30 instance per class label at a time for investigation.

For Classification task, we use RBF classifier [38] for its generalization capability and being faster in training the model. We set the following parameters appropriately for our experiments. The ridge parameter is used to penalize the size of the weights in the output layer. The number of basis functions can also be specified. Noting that large numbers of basis functions produce long training times, we use a single one for our purpose. To improve speed, an approximate version of the logistic function is used as the activation function in the output layer. Here, a single core is used and the Data is split into batches (100 per batch) . Multiple cores may be considered for big datasets. While performing this, Nominal attributes are converted to corresponding binary values and missing values are replaced globally .

## 6. Results and Discussions

This section highlights the experimental results obtained from the proposed approach and compare with the related work available in the literature and makes a substantiate analysis with statistical significance test to understand the suitability of the proposed model in predicting software defects in a most successful way. Table 4 shows our obtained result in terms of six performance measures such as: % accuracy, Precision, Recall, F-score, AUC(Area under the Curve)/ROC value(Receiver operating characteristics) and time to build the model, for .

From Table 5, it is seen that ours method performs best in comparison to NB+MICHAC, RF+MICHAC and RIPPER+MICHAC in terms of having more Precision, Recall, F-score and AUC with 93.5%, 93.5%, 93.6% and 96.2% respectively.

Next, while comparing with others related work using NASA datasets, it can be observed From Table 6 that ours proposed approach outperform all others with accuracy of 98.33% for KC1 dataset, 100% for KC2, JM1 and PC1 datasets. The second best accuracy are from ADBBO-RBF with 84.96% for KC1, SVM for KC2 with 87.12%, 85.36% for JM1 in [45], 95.87% for PC1 in Hybrid SOM and 90% for MW1 in [46].

From Table 7 , we perform Balance comparison amongst all the classifiers and found that our proposed approach outperforms all for all the six datasets mentioned with 0.989,1,1,1,1,0.794 for KC1, KC2, PC1, JM1, MW1 and CM1 datasets, which justifies that our approach has adequately addressed the research question to deal with class imbalance nature of software prediction datasets. The next best for the balance comes from : Rehman et al. [51] for KC1, PC1, JM1 and MW1 with 0.858, 0.971, 0.858 and 0.797 respectively; KNN+ICML for KC2 and CM1with a value of 0.605 and 0.396 respectively.

Table 8 shows a comparative analysis for our proposed approach with JEDIT 4.2 and 4.3 datasets with LR, SVM and RF in terms of Average accuracy, precision and recall values. From the comparisons, it is observed that for both the datasets, our proposed approach is the most accurate with highest precision and recall values with 0.756,0.72 and 0.72 in case of JEDIT 4.2 and 0.8167,0.815 and 0.815 for JEDIT 4.3 respectively in comparison to all others. The second best comes from LR for JEDIT 4.2 for accuracy, Precision and recall with 0.74,0.67 and 0.62 respectively. For JEDIT 4.3, the LR gives best accuracy and precision with 0.79 and 0.65 but for recall RF is the second best.

From Table 9, after comparison in terms of Accuracy (ACC) and Area under the curve (AUC), it is also found that our proposed approach outperforms all others with 99.8%/98.33, 100/100 and 100/100 respectively for KC1, PC1 and JM1 datasets. This proves the effectiveness of our approach to address the another research questions.

From Figure 4, one can understand the level of agreement with each other i.e. short box plot indicates very close agreement where as the comparatively tall box plot indicates their differences in opinion about the aspect or sub-aspect (here it is accuracy).

From Figure 5, Figure 6 , Figure 7 and Figure 8, the result in terms of balance, Accuracy, AUC and F-score justifies the suitability of our proposed approach , Now, we can summarize that the obvious difference in the value of balance, Accuracy, F-score and AUC for all the mentioned approach is worthy of further investigations. Continuing our further investigation on suitability of our proposed approach, we use Win-Draw-Loss (W-D-L) comparison amongst all classifiers, which are illustrated in Table 10, Tabl1 11, Table 12 for NASA Promise dataset in terms of Balance, AUC and Accuracy respectively. The analysis shows that our proposed approach outperform others by becoming winner in all projects where as ADDBBO+RBF stood in second position.

From Table 11 and Table 12, for different projects considered by several authors, it is seen that our proposed approach is the winner in all cases. Further, to extend this experiments to other datasets such as : Softlab AR datasets and Tera promise JEDIT and Mozilla 4 datasets, W-D-L comparisons are provided in Table 13, Table 13 and Table 14 respectively.

The experimental results from Table 13, Table 14 and Table 15 also stand with our proposed approach with all wins in Soft Lab datasets and most number of wins in case of JEDIT and Mozilla 4 datasets. This motivates us to go more detailed analysis with statistical significance test with post-hoc analysis.

**Post-Hoc Statistical significance test**

At first, we perform Kruskal-Wallis rank sum test [62] for multiple classifiers for different performance measures considered by the authors in the respective papers for selected datasets followed by post-hoc Conover and (2) Dunn and (3) Tukey-Kramer (Nemenyi) methods for obtaining the statistical significance test amongst them. Finally, the classifiers which are equally significant are put into Mann Whitney test [63] for comparison. The results obtained along with their corresponding analysis are provided below.

**1. Results of Kruskal-Wallis rank sum test for multiple independent samples among KM(KNN+MCML), Rehman et al. (RE) and Ours(DBBRBF) in terms of Balance**

Kruskal-Wallis chi-squared statistic: 7.384615, degrees of freedom : 2 which is the number of independent samples (or groups), minus 1 and p-value: 0.024914 .

This p-value is for rejection of the omnibus null hypothesis, that all samples (groups) are from the same distribution. The alternate hypothesis that one or more of the independent samples (groups) is different.

The omnibus p-value is at or below the respectable critical threshold of 0.05, so post-hoc pairwise multiple comparison tests are conducted to discern which of the pairs have significantly differences. Three of many possible post-hoc tests are conducted: the methods of (1) Conover, (2) Dunn and (3) Tukey-Kramer (Nemenyi). For the (1) Conover and (2) Dunn methods, the p-value is adjusted according to the family-wide error rate (FWER) procedure of Holm, and then by the false discovery rate (FDR) procedure of Benjamini-Hochberg.

The input data as provided above has ties in the ranks. The (1) Conover and (2) Dunn methods include an adjustment for ties. The method of (3a) Tukey-Kramer (Nemenyi) does not provide for adjustment in the p-values due to ties in the ranks, but when ties are present, the fall-back (3b) Chi-square (Nemenyi) method remains applicable, and its results are also displayed.
Post-hoc p-values of all possible pairs (of samples/ groups) are compactly represented as a lower triangular matrix. Each numerical entry is the p-value of row/column pair, i.e. the null hypotheses that the group represented by a particular column name is different from the group represented by a particular row name. The results obtained are shown in Table 16.
It is worth noting here that, the Conover method tends to show all or most pairs to be significantly different, i.e. its tends to be very liberal. The Dunn and Nemenyi methods seem to be more conservative. Further, the Holm and all other FWER methods of p-value adjustment applied as a secondary step are considered to be liberal relative to the rigorous Benjamnyi-Hochberg FDR method. The results obtained are shown in Table 17.
There are ties revealed in the ranks. The Tukey-Kramer (Nemenyi) method does not have adjustment for ties, and may not be accurate. However, the fall-back Chi-square (Nemenyi) method is still valid, and its p-value results are shown in Table 17.

## 2. Results of Kruskal-Wallis rank sum test for multiple independent samples (KNN+MCML, ADB(ADBBO+RBFNN) and ours

Kruskal-Wallis chi-squared statistic: 7.260504, degrees of freedom: 2, p-value: 0.026510. This p-value is for rejection of the omnibus null hypothesis, that all samples (groups) are from the same distribution. The alternate hypothesis that one or more of the independent samples (groups) is different.
The omnibus p-value is at or below the respectable critical threshold of 0.05, so post-hoc pairwise multiple comparison tests are conducted to discern which of the pairs have significantly differences. Three of many possible post-hoc tests are conducted: the methods of (1) Conover, (2) Dunn and (3) Tukey-Kramer (Nemenyi). For the (1) Conover and (2) Dunn methods, the p-value is adjusted according to the family-wide error rate (FWER) procedure of Holm, and then by the false discovery rate (FDR) procedure of Benjamini-Hochberg.
The input data as provided above has ties in the ranks. The (1) Conover and (2) Dunn methods include an adjustment for ties. The method of (3a) Tukey-Kramer (Nemenyi) does not provide for adjustment in the p-values due to ties in the ranks, but when ties are present, the fall-back (3b) Chi-square (Nemenyi) method remains applicable, and its results are also displayed.
Post-hoc p-values of all possible pairs (of samples/ groups) are compactly represented as a lower triangular matrix. Each numerical entry is the p-value of row/column pair, i.e. the null hypotheses that the group represented by a particular column name is different from the group represented by a particular row name. The results obtained are shown in Table 18.
The Conover method tends to show all or most pairs to be significantly different, i.e. its tends to be very liberal. The Dunn and Nemenyi methods seem to be more conservative. Further, the Holm and all other FWER methods of p-value adjustment applied as a secondary step are considered to be liberal relative to the rigorous Benjamnyi-Hochberg FDR method. The results obtained are shown in Table 19.

There are ties revealed in the ranks. The Tukey-Kramer (Nemenyi) method does not have adjustment for ties, and may not be accurate. However, the fall-back Chi-square (Nemenyi) method is still valid, and its p-value results are shown in Table 19..

### 3. Results of Kruskal-Wallis rank sum test for multiple independent samples (FSOM, SOM, GP, KSA and Ours)

Kruskal-Wallis chi-squared statistic: 10.105516 with degrees of freedom : 4 and p-value: 0.038687. This p-value is for rejection of the omnibus null hypothesis, that all samples (groups) are from the same distribution. The alternate hypothesis that one or more of the independent samples (groups) is different.
The omnibus p-value is at or below the respectable critical threshold of 0.05, so post-hoc pairwise multiple comparison tests are conducted to discern which of the pairs have significantly differences. Three of many possible post-hoc tests are conducted: the methods of (1) Conover, (2) Dunn and (3) Tukey-Kramer (Nemenyi). For the (1) Conover and (2) Dunn methods, the p-value is adjusted according to the family-wide error rate (FWER) procedure of Holm, and then by the false discovery rate (FDR) procedure of Benjamini-Hochberg.
The input data as provided above has ties in the ranks. The (1) Conover and (2) Dunn methods include an adjustment for ties. The method of (3a) Tukey-Kramer (Nemenyi) does not provide for adjustment in the p-values due to ties in the ranks, but when ties are present, the fall-back (3b) Chi-square (Nemenyi) method remains applicable, and its results are also displayed.
Post-hoc p-values of all possible pairs (of samples/ groups) are compactly represented as a lower triangular matrix. Each numerical entry is the p-value of row/column pair, i.e. the null hypotheses that the group represented by a particular column name is different from the group represented by a particular row name. The results obtained are shown in Table 20.
The Conover method tends to show all or most pairs to be significantly different, i.e. its tends to be very liberal. The Dunn and Nemenyi methods seem to be more conservative. Further, the Holm and all other FWER methods of p-value adjustment applied as a secondary step are considered to be liberal relative to the rigorous Benjamnyi-Hochberg FDR method. The results are shown in Table 21.
There are ties revealed in the ranks. The Tukey-Kramer (Nemenyi) method does not have adjustment for ties, and may not be accurate. However, the fall-back Chi-square (Nemenyi) method is still valid, and its p-value results are shown in Table 22.

### 4. Kruskal-Wallis rank sum test for multiple independent samples among LR, SVM, RF and ours

Kruskal-Wallis chi-squared statistic: 9.121637, degrees of freedom : 3 with p-value: 0.027717. This p-value is for rejection of the omnibus null hypothesis, that all samples (groups) are from the same distribution. The alternate hypothesis that one or more of the independent samples (groups) is different.
The omnibus p-value is at or below the respectable critical threshold of 0.05, so post-hoc pairwise multiple comparison tests are conducted to discern which of the pairs have significantly differences. Three of many possible post-hoc tests are conducted: the methods of (1) Conover, (2) Dunn and (3) Tukey-Kramer (Nemenyi). For the (1) Conover and (2) Dunn methods, the p-value is adjusted according to the family-wide error rate (FWER) procedure of Holm, and then by the false discovery rate (FDR) procedure of Benjamini-Hochberg.
The input data as provided above has ties in the ranks. The (1) Conover and (2) Dunn methods include an adjustment for ties. The method of (3a) Tukey-Kramer (Nemenyi) does not provide for adjustment in the p-values due to ties in the ranks, but when ties are present, the fall-back (3b) Chi-square (Nemenyi) method remains applicable, and its results are also displayed.
Post-hoc p-values of all possible pairs (of samples/ groups) are compactly represented as a lower triangular matrix. Each numerical entry is the p-value of row/column pair, i.e. the null hypotheses that the group represented by a particular column name is different from the group represented by a particular row name. The results are shown in Table 23.
The Conover method tends to show all or most pairs to be significantly different, i.e. its tends to be very liberal. The Dunn and Nemenyi methods seem to be more conservative. Further, the Holm and all other FWER methods of p-value adjustment applied as a secondary step are considered to be liberal relative to the rigorous Benjamnyi-Hochberg FDR method. The results are shown in Table 24.
There are ties revealed in the ranks. The Tukey-Kramer (Nemenyi) method does not have adjustment for ties, and may not be accurate. However, the fall-back Chi-square (Nemenyi) method is still valid, and its p-value results are shown in Table 24.

**5, Mann-Whitney U Test Calculator between ours and ADBBO - RBFNN classifier**

The U-value is 0. The critical value of U at $p < .05$ is 2. Therefore, the result is significant at $p < .05$. The Z-Score is -2.50672. The p-value is .01208. The result is significant at $p < .05$.

From all these statistical significance test, it is concluded that ours proposed DBBRBF classifier is most significant classifier in comparison to all others compared and presents its efficacy in predicting the software defects in most suitable.

## 7. Conclusions and Future scope of Research

The paper proposed an efficient distributed based balance supervised instance selection process combining with Radial basis function neural network to predict software defects in a most accurate way. The paper extensively reviewed the related work in this area of research and substantive experiments are conducted to compare the results obtained from the proposed approach with the available related literature and concluded that our proposed one defeats all other in terms different performance measures like: accuracy, F-score, Balance, Recall etc. The box plot representation of all the classifiers used for investigations also opines the suitability of our proposed approach. It shows that our proposed approach deals with data imbalance problem efficiently. Finally, post-hoc Kruskal-Wallis statistical significance test for multiple classifiers are conducted for getting statistical significant ones. To conclude, Mann-Whitney U-test is performed on the best ones to check their solidarity in software defect prediction approaches, After all investigation, it is concluded that ours proposed DBBRBF classifier being a faster classifier outperforms all others in terms accuracy, Balance, F-measure and Recall. In future, the aim shall be to apply the proposed ones to checks more threats to validity with cross project platforms with more performance measures.

Table 1: NASA MDP dataset details [31]

| Name | Language | Features | Instances | Recorded Values | % Defective Instances |
|---|---|---|---|---|---|
| CM1 | C | 40 | 505 | 20200 | 10 |
| JM1 | C | 21 | 10878 | 228438 | 19 |
| KC1 | C++ | 21 | 2107 | 44247 | 15 |
| KC3 | Java | 40 | 458 | 18320 | 9 |
| KC4 | Perl | 40 | 125 | 5000 | 49 |
| MC1 | C and C++ | 39 | 9466 | 369174 | 0.7 |
| MC2 | C | 40 | 161 | 6440 | 32 |
| MW1 | C | 40 | 403 | 16120 | 8 |
| PC1 | C | 40 | 1107 | 44280 | 7 |
| PC2 | C | 40 | 5589 | 223560 | 0.4 |
| PC3 | C | 40 | 1563 | 62520 | 10 |
| PC4 | C | 40 | 1458 | 58320 | 12 |
| PC5 | C++ | 39 | 17186 | 670254 | 3 |
| CM1 | C | 22 | 498 | -- | 9.83 |

Table 2: Characteristics of Softlab dataset [32, 33]

| Project | Number of defective instances | Total number of instances | Number of metrics | % defective instances |
|---|---|---|---|---|
| AR4 | 20 | 107 | 29 | 14.9 |
| AR5 | 8 | 36 | 29 | 22.22 |
| AR6 | 15 | 101 | 29 | 18.7 |

Table 3: JEDIT defect prediction dataset characteristics[34]

| Project | Defect prone modules | Modules | Defect prone ratio |
|---|---|---|---|
| jEdit-4.2 [34] | 48 | 367 | 13.08% |
| jEdit-4.3 [34] | 11 | 492 | 2% |

Input: $D_h$ as training set, C as class variable
|V| = r as the number of terms/features,
b: denotes number of new instances to sample per class,
Output: new $D_h$ represents whole new and fully balanced training set

/* learning phase */
Step-1: for each class $c_k \in C$ do
Step-2: for each term or feature $t_i$, i = 1,…, r do
Step-3: learn probability distribution $P_{ik}$ from $D_h^{\downarrow t_i, C}$

/* sampling phase */
Step-4: new Dh←0;
Step-5: for each class $c_k \in C$ do
Step-6: for p=1 to b do
Step-7: newDoc = newdouble[r+1]
Step-8: for each term or feature $t_i$, i = 1,…, r do
Step-9: newDoc[i] = sample value from $P_{ik}$
Step-10: newDoc[r+1]=$c_k$ //add class label
Step-11: new$D_h$ = new$D_h$ U newDoc
Step-12: return newDh

**Fig. 1. Distribution-based balancing algorithm**

**RBFNN Classifier**

```
trainRBF( in, out, width, MaxError, data )
{
hidden = 0;
net = initRBFNetwork( in, out, hidden ); // init network nodes
do {        // find the data vector that produces the largest error
i = findMaxNetworkError( data, net ); // i = index of vector
addRBFNeuron( net, width, data(i) ); // add neuron to the RBF layer at same
                        point as the above data vector;data(i) = center point
NetError = trainOutputWeights( net, data ); // find overall network error
}
while( NetError > MaxError );
}
```

**Fig 2: Pseudo code of RBFNN Classifier**

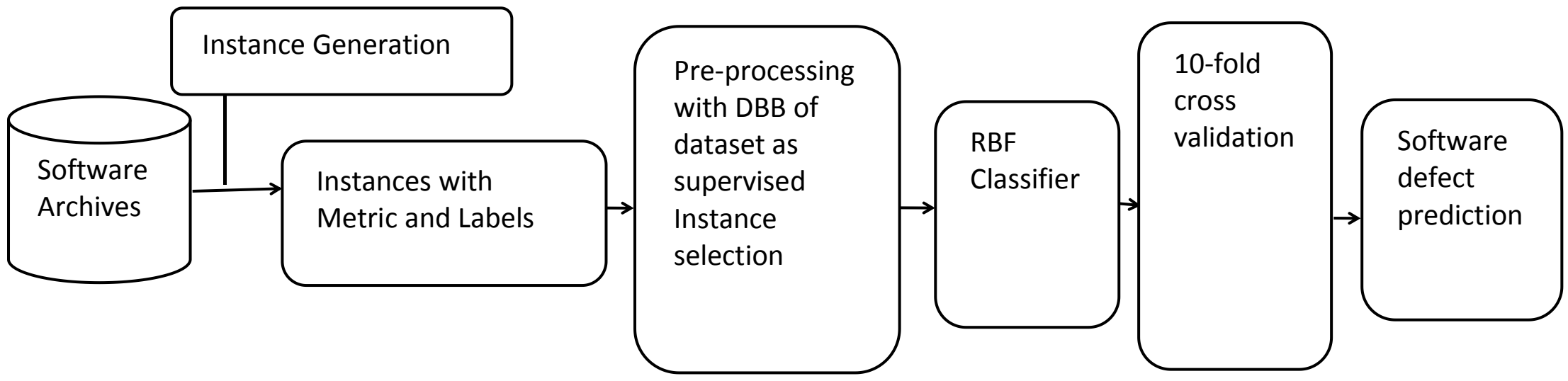

**Figure 3: A common software defect prediction process [39]**

Table 4 : Results of our proposed approach using DBBRBF

| Dataset | % Accuracy | Precision | Recall | F-score | Time in seconds | ROC Area (AUC) |
|---|---|---|---|---|---|---|
| AR4 | 98.34 | 0.985 | 0.985 | 0.984 | 0.01 | 0.999 |
| AR5 | 100 | 1 | 1 | 1 | 0.02 | 1 |
| AR6 | 98.33 | 0.985 | 0.985 | 0.984 | 0.03 | 1 |
| JEDIT 4.2 | 71.6 | 0.72 | 0.72 | 0.72 | 0.03 | 0.803 |
| JEDIT 4.3 | 81.67 | 0.815 | 0.815 | 0.82 | 0.01 | 0.887 |
| KC1 | 98.33 | 0.985 | 0.985 | 0.984 | 0.03 | 0.998 |
| KC2 | 100 | 1 | 1 | 1 | 0.03 | 1 |
| PC1 | 100 | 1 | 1 | 1 | 0.02 | 1 |
| JM1 | 100 | 1 | 1 | 1 | 0.03 | 1 |
| MW1 | 100 | 1 | 1 | 1 | 0.03 | 1 |
| CM1 | 91.67 | 0.916 | 0.916 | 0.916 | 0.12 | 0.983 |
| Mozilla4 | 80 | 0.8 | 0.8 | 0.8 | 0.01 | 0.89 |
| Average | 93.479 | 0.935 | 0.935 | 0.936 | 0.022 | 0.962 |

Table 5: Comparison with AUC values on NASA dataset

| Model/Metric | NB+MICHAC [40] | RF+MICHAC [40] | RIPPER+ MICHAC [40] | DBBRBF (ours) |
|---|---|---|---|---|
| Precision | 0.427 | 0.560 | 0.550 | **0.935** |
| Recall | 0.397 | 0.311 | 0.280 | **0.935** |
| F-score | 0.373 | 0.392 | 0.351 | **0.936** |
| AUC | 0.771 | 0.807 | 0.612 | **0.962** |

NB-Naive Bayes, RF-Random Forest

Table 6: Comparison with % accuracy

| Dataset/ Algorithm | KC1 | KC2 | JM1 | PC1 | MW1 |
|---|---|---|---|---|---|
| ADBBO-RBFNN [41] | 84.96 | 79.51 | 77.03 | 86.29 | - |
| NB [42] | 65.7 | 74 | 60.78 | 60 | - |
| RF[42] | 67.99 | 77.81 | 63.97 | 63.98 | - |
| ANN-ABC [42] | 69 | 79 | 61 | 65 | - |
| Hybrid SOM[43] | 78.43 | 85.98 | 72.33 | 95.87 | - |
| SVM[44] | 79.24 | 87.12 | 70.32 | 92.45 | - |
| DBBRBF(ours) | **98.33** | **100** | **100** | **100** | **100``** |
| [45] | 83.29 | - | 85.36 | - | - |
| [46] | - | - | 81.2 | 90.6 | 90 |
| | | | | | |

Table 7: Balance Comaprison with existing research

| Dataset | DBBRBF (ours) | KNN with ICM [47] | KNN with MCML [47] | Khan et al [48] | Song et al [49] | Jing et al [50] | Rehman et al [51] | Menzies et al [52] | Turhan et al [53] |
|---|---|---|---|---|---|---|---|---|---|
| KC1 | **0.989** | 0.518 | 0.566 | - | 0.707 | 0.705 | 0.858 | - | - |
| KC2 | **1** | 0.538 | 0.605 | - | - | - | - | - | - |
| PC1 | **1** | 0.318 | 0.566 | 0.76 | 0.668 | 0.772 | 0.971 | 0.62 | 0.67 |
| JM1 | **1** | 0.506 | 0.601 | - | 0.585 | 0.664 | 0.858 | - | - |
| MW1 | **1** | - | - | 0.71 | 0.661 | 0.769 | 0.797 | 0.64 | 0.66 |
| CM1 | **0.794** | 0.323 | 0.396 | - | - | - | - | - | - |

Table 8: 10-fold cross validation comparative analysis

| System / Dataset | LR [54] | | | SVM[54] | | | RF[54] | | | DBBRBF (ours) | | |
|---|---|---|---|---|---|---|---|---|---|---|---|---|
| | ACC | Pre | Rec | ACC | Pre | Rec | ACC | Pre | Rec | ACC | Pre | Rec |
| | Avg | Avg | Avg | Avg | Avg | Avg | Avg | Avg | Avg | Avg | Avg | Avg |
| JEDIT 4.2 | 0.74 | 0.67 | 0.62 | 0.71 | 0.59 | 0.57 | 0.62 | 0.56 | 0.60 | **0.756** | **0.72** | **0.72** |
| JEDIT 4.3 | 0.79 | 0.65 | 0.59 | 0.66 | 0.58 | 0.52 | 0.65 | 0.53 | 0.63 | **0.8167** | **0.815** | **0.815** |

LR-Logistic Regression, SVM - Support vector machine and RF-Random Forest

Table 9: ACC and AUC comparison with the results

| Algo/Dataset | KC1 | | PC1 | | JM1 | |
|---|---|---|---|---|---|---|
| | AUC | ACC | AUC | ACC | AUC | ACC |
| DSMOPSO-D +K.Means [55] | 81.58 | 84.33 | 80.47 | 92.13 | 75.2 | 80.44 |
| MOPSO-N [55] | 80.24 | 84.81 | 78.84 | 93.15 | 71.81 | 80.90 |
| Neural Net [55] | 78.98 | 85.79 | 77.75 | 93.44 | 71.24 | 80.94 |
| Bayesian Net [55] | 79.49 | 75.43 | 78.2 | 92.15 | 71.62 | 75.77 |
| Naive Bayes [55] | 79.67 | 82.4 | 75.56 | 89.17 | 66.91 | 80.39 |
| SVM [55] | 50.58 | 83.83 | 50 | 93.05 | 52.7 | 80.82 |
| C4.5 [55] | 67.95 | 84.06 | 67.82 | 93.59 | 66.63 | 80.96 |
| RIPPER [55] | 57.12 | 84.19 | 65.32 | 93.05 | 56.46 | 81.21 |
| Ours DBBRBF | **99.8** | **98.33** | **100** | **100** | **100** | **100** |

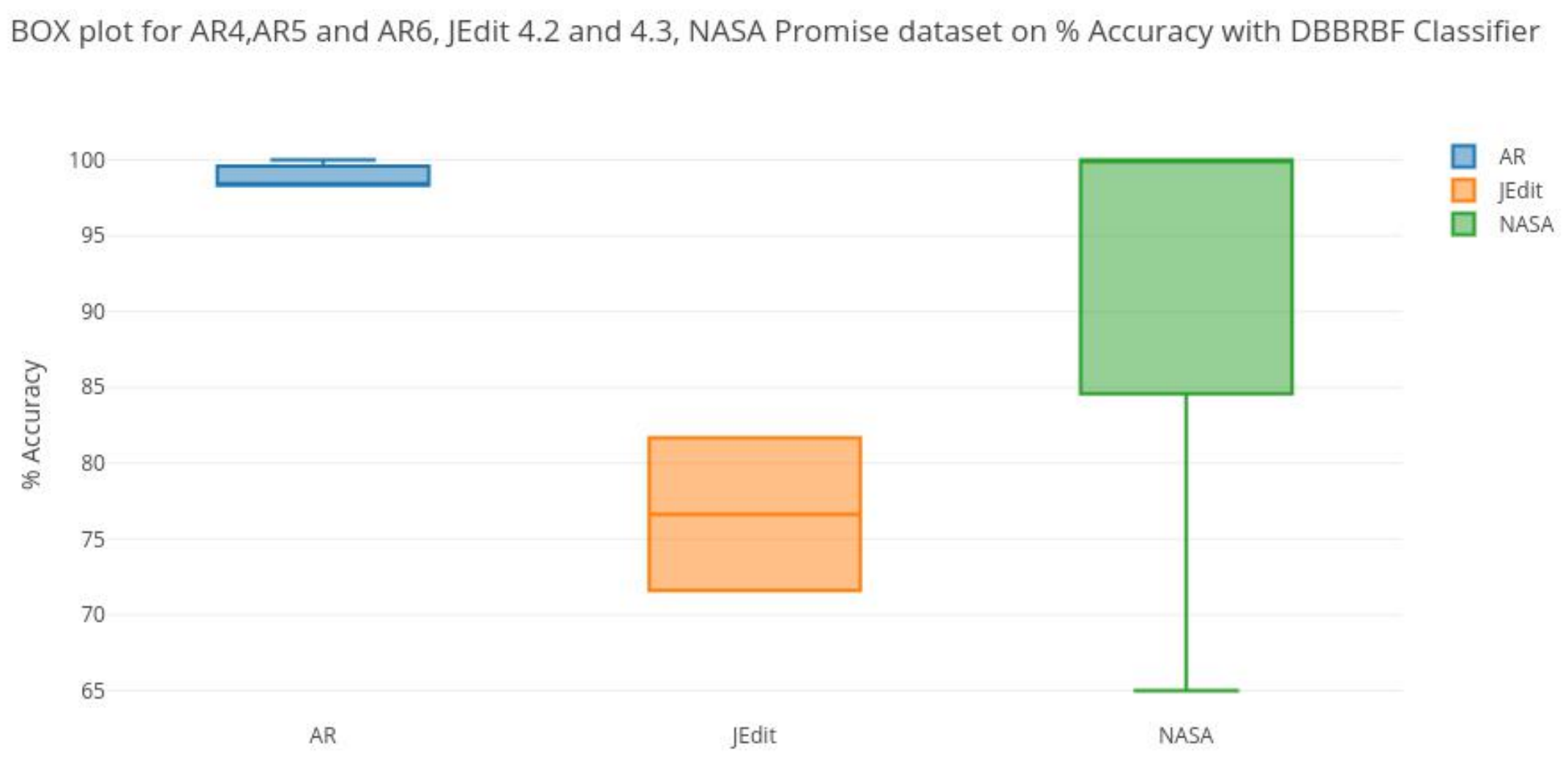

Fig 4: BOX plot for accuracy on software defect prediction datasets used

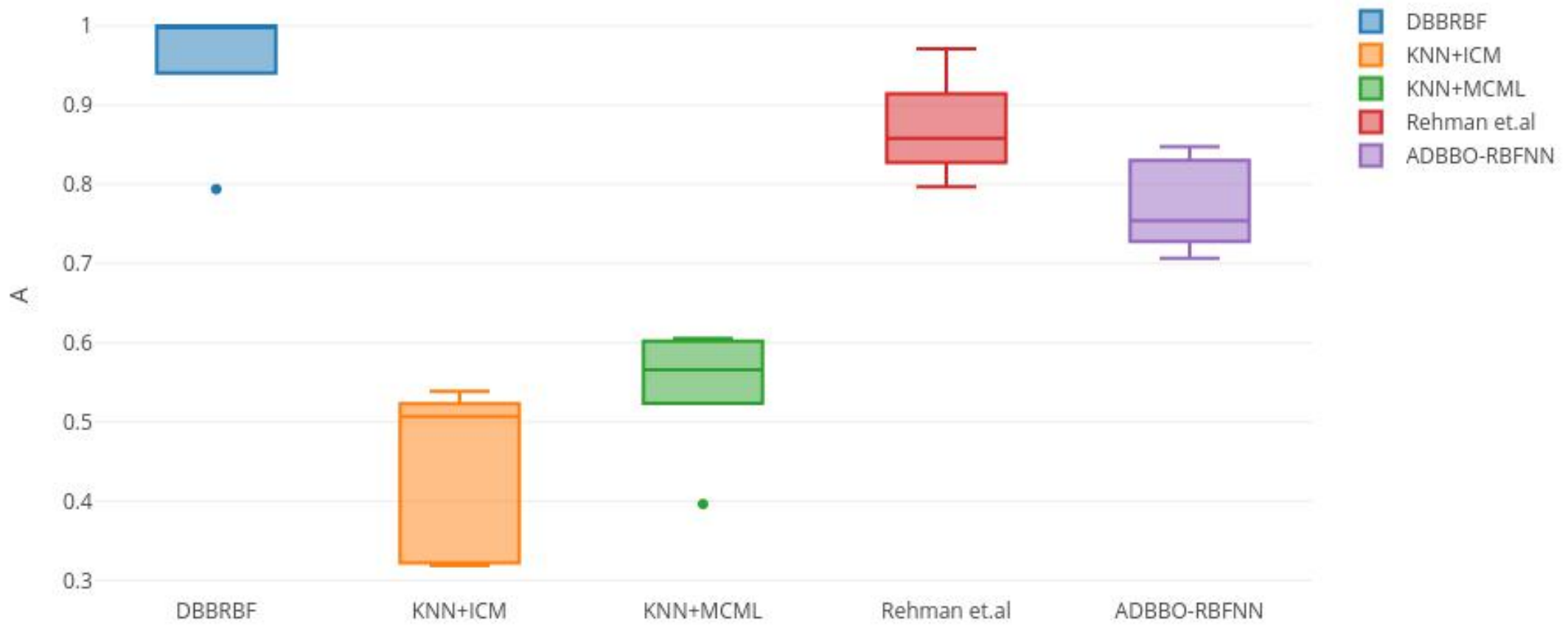

Fig 5: Balance comparison for NASA dataset

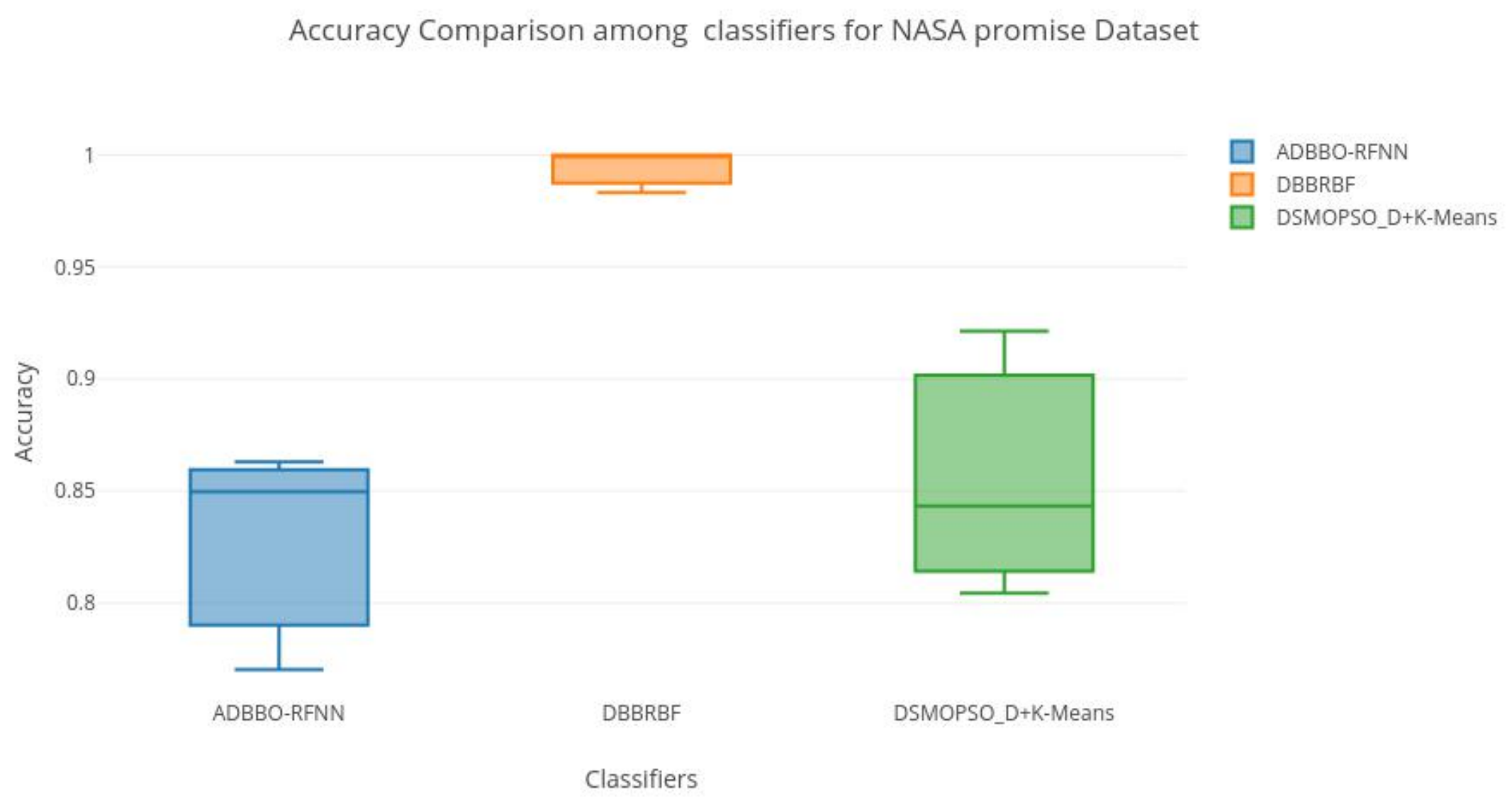

Fig 6: Accuracy comparison on NASA dataset

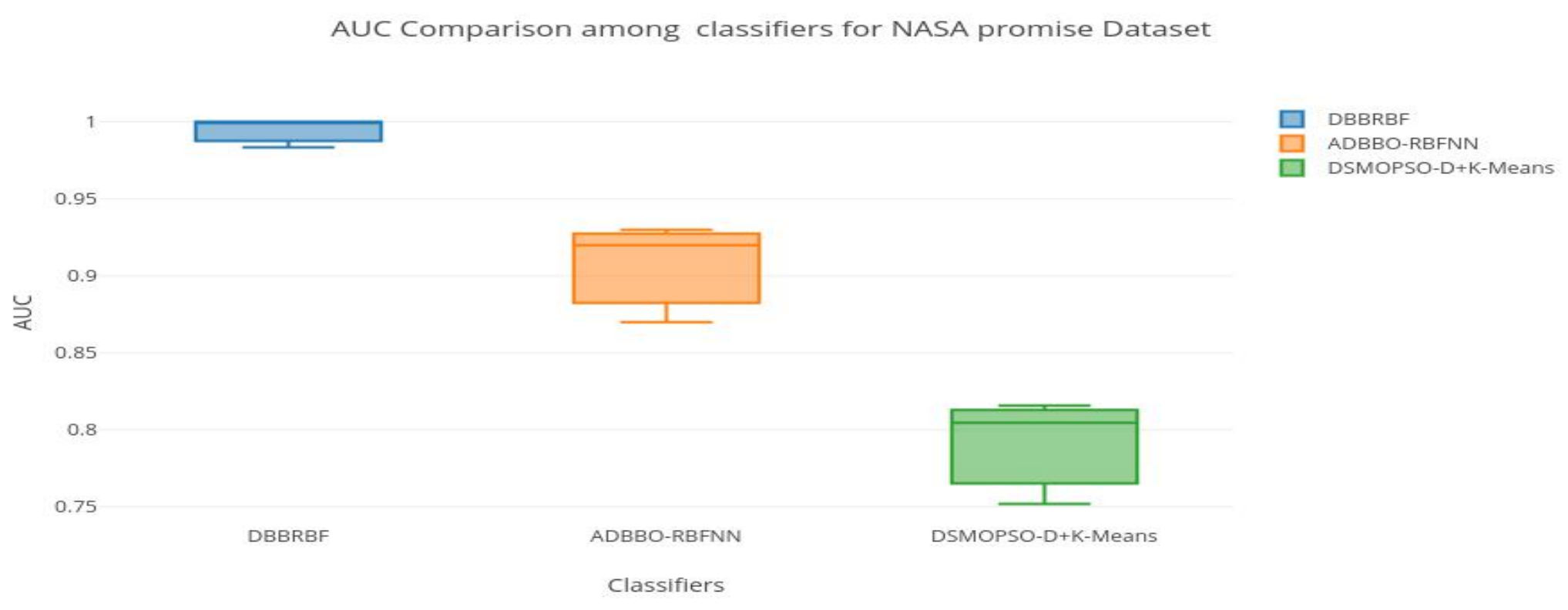

Fig 7: AUC comparison on NASA dataset

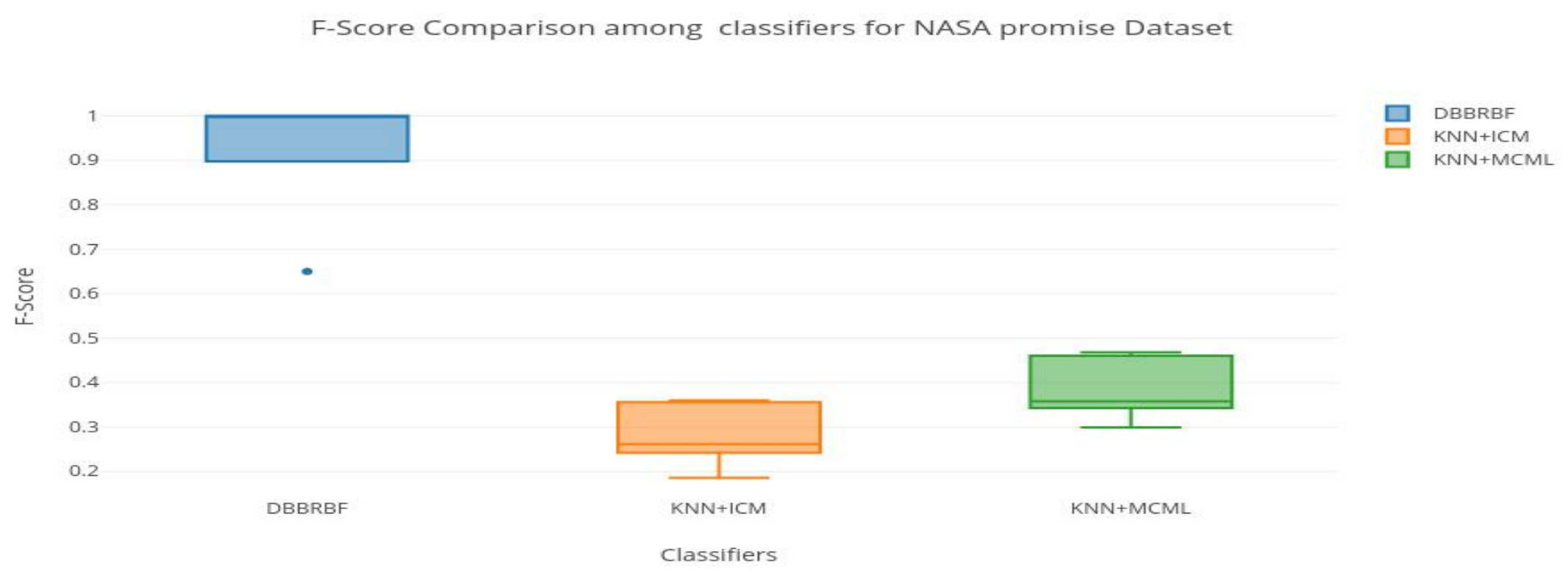

Fig 8: F-score comparison for NASA dataset

Table 10: W/D/L analysis for Balance in NASA promise dataset

| Algorithm | Ours DBBRBF | KNN +ICM | KNN+MCML | ADBBO+RB FNN |
|---|---|---|---|---|
| Ours | ---- | **5/0/0** | **5/0/0** | **5/0/0** |
| KNN+ICM [47] | 0/0/5 | ---- | 0/0/5 | 0/0/5 |
| KNN+MCML [47] | 0/0/5 | 5/0/0 | ----- | 0/0/5 |
| ADBBO+RB FNN [41] | 0/0/5 | 5/0/0 | 5/0/0 | ----- |

Table 11: W-D-L comparison for AUC for NASA Promise dataset

| Algorithm | Ours DBBRBF | DSMOPSO -D+ K-Means | MOPSO-N | Neural net | ADBBO+ RBFNN |
|---|---|---|---|---|---|
| Ours | --- | **3/0/0** | **3/0/0** | **3/0/0** | **5/0/0** |
| DSMOPSO -D+ K-Means [55] | 0/0/3 | ---- | 3/0/0 | 3/0/0 | 0/0/3 |
| MOPSO-N [55] | 0/0/3 | 0/0/3 | ---- | 3/0/0 | 0/0/3 |
| Neural net [55] | 0/0/3 | 0/0/3 | 0/0/3 | ---- | 0/0/3 |
| ADBBO+R BFNN [41] | 0/0/5 | 3/0/0 | 3/0/0 | 3/0/0 | --- |

Table 12: W-D-L comparison for ACCURACY for NASA Promise dataset

| Algorithm | Ours DBBRBF | DSMOPSO _D+ K-Means | MOPSO-N | Neural net | ADBBO+R BFNN |
|---|---|---|---|---|---|
| Ours | --- | **3/0/0** | **3/0/0** | **3/0/0** | **5/0/0** |
| DSMOPSO -D+ K-Means [55] | 0/0/3 | ---- | 3/0/0 | 3/0/0 | 0/0/3 |
| MOPSO-N [55] | 0/0/3 | 0/0/3 | ---- | 3/0/0 | 0/0/3 |
| Neural net [55] | 0/0/3 | 0/0/3 | 0/0/3 | ---- | 0/0/3 |
| ADBBO+R BFNN [41] | 0/0/5 | 3/0/0 | 3/0/0 | 3/0/0 | --- |

Table 13: W-D-L comparison on AUC on SOFTLAB dataset

| Dataset/ Algorithm | Ours DBBRBF | FSOM [56] | SOM [57] | Genetic programming [58] | WDP.KSA analyzer [59] |
|---|---|---|---|---|---|
| AR4 | **4/0/0** | 2/0/2 | 1/0/3 | 0/0/4 | 3/0/1 |
| AR5 | **4/0/0** | 3/0/1 | 2/0/2 | 0/0/4 | 1/0/3 |
| AR6 | **4/0/0** | 3/0/1 | 2/0/2 | 0/0/4 | 1/0/3 |

Table 14: W-D-L comparison on Accuracy on JEDIT dataset

| Algorithm | Ours DBBRBF | LR [54 ] | SVM [54 ] | RF [54 ] |
|---|---|---|---|---|
| JEDIT 4.2 | **2/0/1** | 3/0/0 | 1/0/2 | 0/0/3 |
| JEDIT 4.3 | **3/0/0** | 2/0/1 | 1/0/2 | 0/0/3 |

Table 15: W-D-L comparison on Recall on Mozilla4 dataset

| Algorithm | Ours DBBRBF | DEVREC [60] | Tf-IDF [61] | DLM [61] |
|---|---|---|---|---|
| Mozilla 4 | **2/0/1** | 0/0/3 | 3/0/0 | 1/0/2 |

**Table 16: Results of Post hoc significance test-1 (part-1)**

Conover p-values, further adjusted by the Holm FWER method:

| | KM | RE |
|---|---|---|
| RE | 0.010848 | |
| our | 0.000440 | 0.010848 |

Conover p-values, further adjusted by the Benjamini-Hochberg FDR method: with All possible pairs of null hypotheses are being compared below.

| | KM | RE |
|---|---|---|
| RE | 0.005424 | |
| our | 0.000440 | 0.005424 |

Conover p-values, without p-value adjustment:

| | KM | RE |
|---|---|---|
| RE | 0.005424 | |
| our | 0.000147 | 0.005424 |

**Table 17: Results of Post hoc significance test-1 (part-2)**

Dunn p-values, further adjusted by the Holm FWER method:

| | KM | RE |
|---|---|---|
| RE | 0.348463 | |
| our | 0.019735 | 0.348463 |

Dunn p-values, further adjusted by the Benjamini-Hochberg FDR method: with All possible pairs of null hypotheses are being compared above.

| | KM | RE |
|---|---|---|
| RE | 0.174231 | |
| our | 0.019735 | 0.174231 |

Dunn p-values, without p-value adjustment:

| | KM | RE |
|---|---|---|
| RE | 0.174231 | |
| our | 0.006578 | 0.174231 |

Tukey-Kramer (Nemenyi) p-values, without p-value adjustment:

| | KM | RE |
|---|---|---|
| RE | 0.372059 | |
| our | 0.019945 | 0.372059 |

Chi-square (Nemenyi) p-values, without p-value adjustment:

| | KM | RE |
|---|---|---|
| RE | 0.397295 | |
| our | 0.024914 | 0.397295 |

**Table 18: Results of Post hoc significance test-2 (part-1)**

Conover p-values, further adjusted by the Holm FWER method:

| | ADB | KM |
|---|---|---|
| KM | 0.017165 | |
| ours | 0.017165 | 0.000768 |

Conover p-values, further adjusted by the Benjamini-Hochberg FDR method: with All possible pairs of null hypotheses are being compared below.

| | ADB | KM |
|---|---|---|
| KM | 0.008583 | |
| ours | 0.008583 | 0.000768 |

Conover p-values, without p-value adjustment:

| | ADB | KM |
|---|---|---|
| KM | 0.008583 | |
| ours | 0.008583 | 0.000256 |

**Table 19: Results of Post hoc significance test-2 (part-2)**

Dunn p-values, further adjusted by the Holm FWER method:

| | ADB | KM |
|---|---|---|
| KM | 0.355789 | |
| ours | 0.355789 | 0.021146 |

Dunn p-values, further adjusted by the Benjamini-Hochberg FDR method: with All possible pairs of null hypotheses are being compared above.

| | ADB | KM |
|---|---|---|
| KM | 0.177895 | |
| ours | 0.177895 | 0.021146 |

Dunn p-values, without p-value adjustment:

| | ADB | KM |
|---|---|---|
| KM | 0.177895 | |
| ours | 0.177895 | 0.007049 |

Tukey-Kramer (Nemenyi) p-values, without p-value adjustment:

| | ADB | KM |
|---|---|---|
| KM | 0.372059 | |
| ours | 0.372059 | 0.019945 |

Chi-square (Nemenyi) p-values, without p-value adjustment:

| | ADB | KM |
|---|---|---|
| KM | 0.403506 | |
| ours | 0.403506 | 0.02651 |

**Table 20: Results of Post hoc significance test-3 (part-1)**

Conover p-values, further adjusted by the Holm FWER method:

| | FSOM | GP | KSA | SOM |
|---|---|---|---|---|
| GP | 0.134000 | | | |
| KSA | 1.000000 | 0.427105 | | |
| SOM | 1.000000 | 0.401491 | 1.000000 | |
| ours | 0.013265 | 0.549663 | 0.044063 | 0.038846 |

Conover p-values, further adjusted by the Benjamini-Hochberg FDR method: with all possible pairs of null hypotheses are being compared below.

| | FSOM | GP | KSA | SOM |
|---|---|---|---|---|
| GP | 0.047857 | | | |
| KSA | 0.498856 | 0.142368 | | |
| SOM | 0.533100 | 0.133830 | 0.886203 | |
| ours | 0.013265 | 0.196308 | 0.018359 | 0.018359 |

Conover p-values, without p-value adjustment:

| | FSOM | GP | KSA | SOM |
|---|---|---|---|---|
| GP | 0.019143 | | | |
| KSA | 0.399085 | 0.085421 | | |
| SOM | 0.479790 | 0.066915 | 0.886203 | |
| ours | 0.001326 | 0.137416 | 0.005508 | 0.004316 |

**Table 21: Results of Post hoc significance test-3 (part-2)**

Dunn p-values, further adjusted by the Holm FWER method:

| | FSOM | GP | KSA | SOM |
|---|---|---|---|---|
| GP | 0.572175 | | | |
| KSA | 1.000000 | 1.0 | | |
| SOM | 1.000000 | 1.0 | 1.000000 | |
| ours | 0.059878 | 1.0 | 0.223162 | 0.197998 |

Dunn p-values, further adjusted by the Benjamini-Hochberg FDR method: with All possible pairs of null hypotheses are being compared above.

| | FSOM | GP | KSA | SOM |
|---|---|---|---|---|
| GP | 0.204348 | | | |
| KSA | 0.718777 | 0.389427 | | |
| SOM | 0.718777 | 0.389427 | 0.927004 | |
| ours | 0.059878 | 0.447954 | 0.092984 | 0.092984 |

Dunn p-values, without p-value adjustment:

| | FSOM | GP | KSA | SOM |
|---|---|---|---|---|
| GP | 0.081739 | | | |
| KSA | 0.582533 | 0.233656 | | |
| SOM | 0.646899 | 0.199629 | 0.927004 | |
| ours | 0.005988 | 0.313568 | 0.027895 | 0.022 |

Tukey-Kramer (Nemenyi) p-values, without p-value adjustment:

| | FSOM | GP | KSA | SOM |
|---|---|---|---|---|
| GP | 0.412662 | | | |
| KSA | 0.982246 | 0.759068 | | |
| SOM | 0.991071 | 0.704809 | 0.999984 | |
| ours | 0.048530 | 0.853584 | 0.183042 | 0.150585 |

**Table 22: Results of Post hoc significance test-3 (part-3)**

Chi-square (Nemenyi) p-values, without p-value adjustment:

| | FSOM | GP | KSA | SOM |
|---|---|---|---|---|
| GP | 0.552822 | | | |
| KSA | 0.989674 | 0.840979 | | |
| SOM | 0.994867 | 0.800667 | 0.999991 | |
| ours | 0.109353 | 0.907423 | 0.304701 | 0.262993 |

**Table 23: Results of Post hoc significance test-4 (part-1)**

Conover p-values, further adjusted by the Holm FWER method:

| | LR | RF | SVM |
|---|---|---|---|
| RF | 0.005852 | | |
| SVM | 0.082242 | 0.113692 | |
| ours | 0.574541 | 0.003439 | 0.044942 |

Conover p-values, further adjusted by the Benjamini-Hochberg FDR method: with All possible pairs of null hypotheses are being compared below.

| | LR | RF | SVM |
|---|---|---|---|
| RF | 0.003511 | | |
| SVM | 0.041121 | 0.068215 | |
| ours | 0.574541 | 0.003439 | 0.022471 |

Conover p-values, without p-value adjustment:

| | LR | RF | SVM |
|---|---|---|---|
| RF | 0.001170 | | |
| SVM | 0.027414 | 0.056846 | |
| ours | 0.574541 | 0.000573 | 0.011236 |

**Table 24: Results of Post hoc significance test-4 (part-2)**

Dunn p-values, further adjusted by the Holm FWER method:

| | LR | RF | SVM |
|---|---|---|---|
| RF | 0.086106 | | |
| SVM | 0.576291 | 0.576291 | |
| ours | 0.776743 | 0.046122 | 0.449176 |

Dunn p-values, further adjusted by the Benjamini-Hochberg FDR method:
with All possible pairs of null hypotheses are being compared above.

| | LR | RF | SVM |
|---|---|---|---|
| RF | 0.051664 | | |
| SVM | 0.288145 | 0.337482 | |
| ours | 0.776743 | 0.046122 | 0.224588 |

Dunn p-values, without p-value adjustment:

| | LR | RF | SVM |
|---|---|---|---|
| RF | 0.017221 | | |
| SVM | 0.192097 | 0.281235 | |
| ours | 0.776743 | 0.007687 | 0.112294 |

Tukey-Kramer (Nemenyi) p-values, without p-value adjustment:

| | LR | RF | SVM |
|---|---|---|---|
| RF | 0.081362 | | |
| SVM | 0.561540 | 0.704423 | |
| ours | 0.992094 | 0.038998 | 0.387064 |

Chi-square (Nemenyi) p-values, without p-value adjustment:

| | **LR** | **RF** | **SVM** |
|---|---|---|---|
| **RF** | 0.128612 | | |
| **SVM** | 0.636609 | 0.762346 | |
| **ours** | 0.994080 | 0.068626 | 0.471393 |